\documentclass[a4paper,conference]{IEEEtran}
\usepackage[table]{xcolor}

\newcommand{\exportFigures}{true} 
\usepackage{tikz}
\usetikzlibrary{backgrounds}
\tikzstyle{load}   = [ultra thick,-latex]
\tikzstyle{stress} = [-latex]
\tikzstyle{dim}    = [latex-latex]
\tikzstyle{axis}   = [-latex,black!55]

\definecolor{green(pigment)}{rgb}{0.0, 0.65, 0.31}
\definecolor{frenchblue}{rgb}{0.0, 0.45, 0.73} 
\definecolor{mediumcandyapplered}{rgb}{0.89, 0.02, 0.17}

\usetikzlibrary{spy}
\usetikzlibrary{backgrounds}
\tikzset{background fill/.style={background rectangle/.style={fill=#1},show background rectangle}}

\usepackage{tkz-euclide}
\usetikzlibrary{arrows}
\usepackage{setspace}
\usepackage{bm}
\usepackage{mathtools}
\usepackage{graphicx}
\usepackage{relsize}
\usepackage{cite}
\usepackage{amsmath}
\usepackage{color}
\usepackage{psfrag}
\usepackage{amssymb}
\usepackage{dblfloatfix}
\usepackage[font=small]{caption}
\usepackage{tabularx}
\usepackage{makecell}
\usepackage{lipsum}
\usepackage{float}
\usepackage{subfiles}
\usepackage{epstopdf}
\usepackage{verbatim}
\usepackage{subfig}
\usepackage{amsmath}
\usepackage{ifthen}
\usepackage{pgfplots}
\usepackage{pgfplotstable}
\usepackage{tikzscale}
\usepackage{acronym}
\usepackage{amsfonts}
\usepackage{tikz-3dplot}
\usetikzlibrary{3d}
\usetikzlibrary{shapes}
\usetikzlibrary {patterns,patterns.meta}
\usepackage{pgffor}
\usetikzlibrary{shapes.geometric}
\usetikzlibrary{angles,quotes}

\renewcommand{\baselinestretch}{1} 

\colorlet{veccol}{green!50!black}
\colorlet{projcol}{blue!70!black}
\colorlet{myblue}{blue!80!black}
\colorlet{myred}{red!90!black}
\colorlet{mydarkblue}{blue!50!black}
\tikzset{>=latex} 
\tikzstyle{proj}=[projcol!80,line width=0.08] 
\tikzstyle{area}=[draw=veccol,fill=veccol!80,fill opacity=0.6]
\tikzstyle{vector}=[-stealth,myblue,thick,line cap=round]
\tikzstyle{unit vector}=[->,veccol,thick,line cap=round]
\tikzstyle{dark unit vector}=[unit vector,veccol!70!black]
\usetikzlibrary{angles,quotes} 

\newcommand{\ist}{\hspace*{.3mm}}
\newcommand{\rmv}{\hspace*{-.3mm}}
\newcommand{\iist}{\hspace*{1mm}}

\newcommand{\nn}{\nonumber}

\providecommand{\norm}[1]{\lVert#1\rVert}

\DeclareMathAlphabet{\mathsfbr}{OT1}{cmss}{m}{n}
\SetMathAlphabet{\mathsfbr}{bold}{OT1}{cmss}{bx}{n}
\DeclareRobustCommand{\msf}[1]{%
	\ifcat\noexpand#1\relax\msfgreek{#1}\else\mathsfbr{#1}\fi
}

\makeatletter
\newcommand{\msfgreek}[1]{\csname s\expandafter\@gobble\string#1\endcsname}
\makeatother

\DeclareRobustCommand{\mcal}[1]{%
	\ifcat\noexpand#1\relax\mathnormal{#1}\else\cal{#1}\fi
}
\DeclareRobustCommand{\BM}[1]{%
	\ifcat\noexpand#1\relax\bm{\boldUppercaseItalicGreek{#1}}\else\bm{#1}\fi
}
\makeatletter
\newcommand{\boldUppercaseItalicGreek}[1]{\csname var\expandafter\@gobble\string#1\endcsname}
\makeatother
\newcommand{\rv}[1]{\msf{#1}}
\newcommand{\RV}[1]{\bm{\msf{#1}}}

\newcommand{\V}[1]{\bm{#1}}
\newcommand{\M}[1]{\BM{#1}}

\DeclareMathVersion{timesmath}
\SetSymbolFont{letters}{timesmath}{OML}{ntxmi}{m}{it}
\DeclareMathVersion{timesmathbold}
\SetSymbolFont{letters}{timesmathbold}{OML}{ntxmi}{b}{it}

\acrodef{2d}[2D]{two-dimensional}
\acrodef{3d}[3D]{three-dimensional}
\acrodef{5g}[5G]{5th-generation}
\acrodef{pnt}[PNT]{positioning, navigation and timing}
\acrodef{pa}[PA]{physical anchor}
\acrodef{bs}[BS]{base station}
\acrodef{awgn}[AWGN]{additive white Gaussian noise}
\acrodef{va}[VA]{virtual anchor}
\acrodef{mpc}[MPC]{multipath component}
\acrodef{fa}[FA]{false alarm}
\acrodef{far}[FAR]{false alarm rate}
\acrodef{pva}[PVA]{potential \ac{va}}
\acrodef{ef}[MF]{map feature}
\acrodef{pef}[PMF]{potential \ac{ef}}
\acrodef{smc}[SMC]{specular multipath component}
\acrodef{psmc}[PSMC]{potential \ac{smc}}
\acrodef{slam}[SLAM]{simultaneous localization and mapping}
\acrodef{pmf}[PMF]{probability mass function}
\acrodef{pdf}[PDF]{probability density function}
\acrodef{cdf}[CDF]{cummulative distribution function}
\acrodef{bp}[BP]{belief propagation}
\acrodef{spa}[SPA]{sum-product algorithm}
\acrodef{mmse}[MMSE]{minimum mean-square error}
\acrodef{simo}[SIMO]{single-input-multiple-output}
\acrodef{mimo}[MIMO]{multiple-input-multiple-output}
\acrodef{mmwave}[mmWave]{millimeter-wave}
\acrodef{ue}[UE]{mobile user}
\acrodef{los}[LoS]{line-of-sight}
\acrodef{aoa}[AoA]{angle-of-arrival}
\acrodef{aod}[AoD]{angle-of-departure}
\acrodef{roi}[RoI]{region-of-interest}
\acrodef{snr}[SNR]{signal-to-noise ratio}
\acrodef{ospa}[OSPA]{optimal subpattern assignment}
\acrodef{mospa}[MOSPA]{mean \ac{ospa}}
\acrodef{dmc}[DMC]{dense multipath component}
\acrodef{sr}[SR]{super-resolution}
\acrodef{sbl}[SBL]{sparse Bayesian learning}
\acrodef{sde}[SDE]{sequential detection and estimation}
\acrodef{kest}[KEST]{Kalman enhanced super resolution tracking}
\acrodef{da}[DA]{data association}
\acrodef{fim}[FIM]{Fisher information matrix}
\acrodef{crlb}[CRLB]{Cram\'{e}r–Rao lower bound}
\acrodef{pcrlb}[PCRLB]{posterior \ac{crlb}}
\acrodef{va}[VA]{virtual anchor}
\acrodef{eadf}[EADF]{effective aperture distribution function}
\acrodef{lhf}[LHF]{likelihood function}
\acrodef{rmse}[RMSE]{root mean square error} 
\acrodef{ps}[PS]{point scatterer}
\acrodef{imm}[IMM]{interacting multiple model}

\newcommand{\tikzfolder}{./compiledPlots/}
\usetikzlibrary{backgrounds}
\usetikzlibrary{calc,arrows.meta}
\ifthenelse{\equal{\exportFigures}{true}}
{
  \usepgfplotslibrary{external}\tikzexternalize[prefix=\tikzfolder]
  \tikzset{external/system call={pdflatex \tikzexternalcheckshellescape -halt-on-error -interaction=batchmode -jobname "\image" "\texsource"}}

}
{}

\def\addlegendimage{\csname pgfplots@addlegendimage\endcsname}
\def\addlegendentry{\csname pgfplots@addlegendentry\endcsname}

\newcommand{\pgfref}[1]
{\ifthenelse{\equal{\exportFigures}{true}}
{\tikzexternaldisable\ref{#1}\tikzexternalenable}
{\ref{#1}}}

\pgfplotsset{compat=newest} 

\newlength\figureheight 
\newlength\figurewidth 
\usetikzlibrary{patterns,arrows}
\usepgfplotslibrary{colormaps}
\usetikzlibrary{plotmarks}
\pgfplotsset{every axis/.append style={
  label style={font=\normalsize},
  tick label style={font=\scriptsize},
  xticklabel={
   \ifdim \tick pt < 0pt
    \pgfmathparse{abs(\tick)}%
    \llap{$-{}$}\pgfmathprintnumber{\pgfmathresult}
   \else
    \pgfmathprintnumber{\tick}
   \fi}
   }}
\usetikzlibrary{decorations.markings}
\makeatletter
\tikzset{
  nomorepostactions/.code={\let\tikz@postactions=\pgfutil@empty},
  mymarkfixednumber/.style n args={3}{decoration={markings,
    mark= between positions 0.1 and 1 step (1/#3)*\pgfdecoratedpathlength with{%
        \tikzset{#2,every mark}\tikz@options
        \pgfuseplotmark{#1}%
      },  
    },
    postaction={decorate},
    /pgfplots/legend image post style={
        mark=#1,mark options={#2},every path/.append style={nomorepostactions}
    },
  },
}
\tikzset{
  nomorepostactions/.code={\let\tikz@postactions=\pgfutil@empty},
  mymarkfixeddistance/.style n args={3}{decoration={markings,
    mark= between positions 0 and 1 step #3cm with{%
        \tikzset{#2,every mark}\tikz@options
        \pgftransformresetnontranslations%
        \pgfuseplotmark{#1}%
      },  
    },
    postaction={decorate},
    /pgfplots/legend image post style={
        mark=#1,mark options={#2},every path/.append style={nomorepostactions}
    },
  },
}
\tikzset{
  nomorepostactions/.code={\let\tikz@postactions=\pgfutil@empty},
  mymark/.style n args={3}{decoration={markings,
    mark= between positions 0 and 1 step 0.75cm with{%
        \tikzset{#2,every mark}\tikz@options
        \pgftransformresetnontranslations%
        \pgfuseplotmark{#1}%
      },  
    },
    postaction={decorate},
    /pgfplots/legend image post style={
        mark=#1,mark options={#2},every path/.append style={nomorepostactions}
    },
  },
}
\makeatother

\definecolor{colA}{rgb}{0,0,1}
\definecolor{colB}{rgb}{1.00000,0.0,0.00000}
\definecolor{colC}{rgb}{0.1333,0.5451,0.1333}
\definecolor{colD}{rgb}{0.9,0.0,0.9}
\definecolor{colE}{rgb}{0,1,1}
\definecolor{colF}{rgb}{1.00000,0.55,0.00000}

\def\mymarksize{1.5}
\def\mymarksize2{2.5}

\pgfdeclareplotmark{markarrow}
{%
  \pgfpathmoveto{\pgfqpoint{\pgfplotmarksize}{0pt}}
  \pgfpathlineto{\pgfqpointpolar{130}{1.5\pgfplotmarksize}}
  \pgfpathmoveto{\pgfqpoint{\pgfplotmarksize}{0pt}}
  \pgfpathlineto{\pgfqpointpolar{-130}{1.5\pgfplotmarksize}}
  \pgfusepathqstroke
}

\IEEEoverridecommandlockouts

\renewcommand{\baselinestretch}{1}

\begin{document}
\allowdisplaybreaks
\frenchspacing

\title{\huge A Belief Propagation Algorithm for Multipath-based SLAM \\ \hspace*{0mm}with Multiple Map Features:\hspace*{-0.3mm} A mmWave MIMO Application}
\author{\IEEEauthorblockN{Xuhong Li$^\ast$, Xuesong Cai$^\ast$, Erik Leitinger$^\dagger$, and Fredrik Tufvesson$^\ast$}
	\IEEEauthorblockA{$^\ast$Department of Electrical and Information Technology, Lund University, Sweden. \\ $^\dagger$Signal Processing and Speech Communication Laboratory, Graz University of Technology, Austria \\ Email: \{xuhong.li, xuesong.cai, fredrik.tufvesson\}@eit.lth.se, erik.leitinger@tugraz.at }
	}

\maketitle

\begin{abstract}
In this paper, we present a multipath-based \ac{slam} algorithm that continuously adapts mulitiple \ac{ef} models describing specularly reflected \acp{mpc} from flat surfaces and point-scattered \acp{mpc}, respectively. We develop a Bayesian model for sequential detection and estimation of interacting \ac{ef} model parameters, \ac{ef} states and mobile agent's state including position and orientation. The Bayesian model is represented by a factor graph enabling the use of \ac{bp} for efficient computation of the marginal posterior distributions. The algorithm also exploits amplitude information enabling reliable detection of ``weak'' \acp{ef} associated with \acp{mpc} of very low \acp{snr}. The performance of the proposed algorithm is evaluated using real \ac{mmwave} \ac{mimo} measurements with single base station setup. Results demonstrate the excellent localization and mapping performance of the proposed algorithm in challenging dynamic outdoor scenarios.

\end{abstract}

\IEEEpeerreviewmaketitle

\section{Introduction}
5G and beyond networks exploiting \ac{mmwave} spectrum and massive \ac{mimo} techniques show great potential in providing exceptional localization and sensing services even in harsh environments like urban canyons and indoors. With increased signal bandwidth and array aperture providing superior spatial resolution, specular \acp{mpc} associated with distinct \acfp{ef} can be fine resolved and therefore exploited for \acf{slam} \cite{Erik_SLAM_TWC2019, Erik_DataFusion2023, RicoJSAC2019, YuGe_mmWaveSLAM_JSAC2022, MohammadmmWaveTVT2023}. Leveraging \acp{mpc} largely improves the localization accuracy and robustness, particularly in environments with strong multipath propagation and \ac{los} obstruction. Moreover, multipath-based \ac{slam} alleviates infrastructure needs, even single-base station localization becomes viable.

\acp{ef} for radio signals mostly refer to \acp{va} denoting the mirror images of \acp{pa} (e.g., base stations) w.r.t., flat surfaces and modeling signal specular reflection. However, the importance of considering diverse \ac{ef} models representing different environment interacting objects such as extended surfaces \cite{Erik_DataFusion2023} and point scatterers \cite{HediehEuCAP2024, GentnerTWC2016, KimGranSveKimWym:TVT2022, YuGe_mmWaveSLAM_JSAC2022} is increasingly recognized. Different types of \acp{ef} often coexist in complex propagation environments and are gradually starting to be considered in multipath-based \ac{slam} approaches, e.g., \cite{GentnerTWC2016} models \acp{va}, \acp{ps} and their combination, \cite{KimGranSveKimWym:TVT2022,YuGe_mmWaveSLAM_JSAC2022} incorporate the modeling and detection of \ac{va}- and \ac{ps}-type of \acp{ef} in a PMBM-based \ac{slam} framework. In general, the unknown \acp{ef} types, the unknown time-varying \acp{ef} number in dynamic scenarios, and the association uncertainty of measurements with \acp{ef} present as major challenges for multipath-based \ac{slam}.

In this paper, we extended a multipath-based \ac{slam} algorithm \cite{Erik_SLAM_TWC2019,LeitingerICC2019} by incorporating different statistical models for \ac{va}- and \ac{ps}-type \acp{ef}, and \ac{mimo} setup. The time-evolution of the interacting multiple \ac{ef} model parameters are described by a discrete Markov chain that is incorporated into the Bayesian model formulating the \ac{slam} problem. Using the \ac{mpc} estimates, i.e., distances, \acp{aoa}, \acp{aod} and amplitudes, from a snapshot-based parametric channel estimator SAGE \cite{Xuesong_EADF2023} as measurements, the proposed \acf{bp} algorithm sequentially adapts the interacting multiple \ac{ef} model parameters along with the detection and estimation of the mobile agent state (including time-varying position and orientation), and the states of \acp{ef}. Furthermore, the algorithm uses the statistics of \ac{mpc} amplitudes to determine the unknown and time varying detection probabilities, which improves the detectability and maintenance of low \ac{snr} \acp{ef}. The performance is validated using real \ac{mmwave} \ac{mimo} measurements in a challenging outdoor dynamic environment with single-\ac{pa} setup.

\section{Geometrical Model of the Environment}
\label{sec:GeometricalRelations}
\begin{figure}[!t]
	\centering
	\includegraphics[]{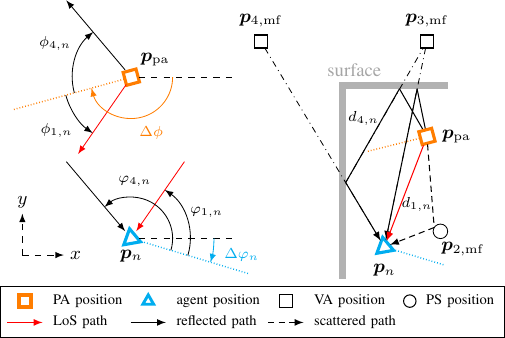}\\[1mm]
	\caption{Geometrical depiction of a \ac{mimo} radio propagation environment, where \acp{mpc} are represented by \acp{ef}, e.g., \acp{va} and \acp{ps}.}	 
	\label{fig:GraphicalOverview}
\end{figure}

We consider a \ac{mmwave} \ac{mimo} system operating in a dynamic scenario.  For the sake of brevity, we assume a two-dimensional scenario with horizontal-only signal propagation. At each discrete time $ n $, a \ac{pa} with known position $\V{p}_{\mathrm{pa}} =[p_{\mathrm{pa},\mathrm{x}} \iist p_{\mathrm{pa},\mathrm{y}}]^{\mathrm{T}}$ transmits a radio signal and a mobile agent at unknown time-varying position $\V{p}_{n} =[p_{n,\mathrm{x}} \iist p_{n,\mathrm{y}}]^{\mathrm{T}}$ acts as a receiver.\footnote{The proposed algorithm can be easily reformulated for the case where the mobile agent acts as a transmitter and the \ac{pa} acts as a receiver.} We assume time and frequency synchronization between the \ac{pa} and the mobile agent. The \ac{pa} is equipped with a $ N_{\mathrm{tx}} $-element antenna array with known orientation $\Delta\phi$, and the mobile agent is equipped with a $ N_{\mathrm{rx}} $-element antenna array with unknown azimuth orientation $\Delta\varphi_{n}$, respectively. The positions $ \V{p}_{\mathrm{pa}} $ and $ \V{p}_{n} $ refer to the center of gravity of the arrays. Specularly reflected \acp{mpc} and scattered \acp{mpc} can be modeled by \acp{va} and \acp{ps}, reflectively. The \ac{pa}, \acp{va} and \acp{ps} are collectively referred to as \acp{ef} at unknown but fixed positions $ \V{p}_{l,\mathrm{mf}} =[p_{l,\mathrm{mf},\mathrm{x}} \iist p_{l,\mathrm{mf},\mathrm{y}}]^{\mathrm{T}}$, with $l \in \{1,\dots,L_{n}\}$. 

As shown in Fig.~\ref{fig:GraphicalOverview}, for \acp{mpc} generated by \ac{va}-type of \acp{ef}, the propagation distances and \acp{aoa} at time $ n $ are given by $ d_{l,n} = \mathrm{d}_{\mathrm{va}}(\V{p}_{n}, \V{p}_{l,\mathrm{mf}})$ and $ \varphi_{l,n} = \angle(\V{p}_{n}, \V{p}_{l,\mathrm{mf}}, \Delta\varphi_{n})$, respectively. For \acp{mpc} originated from \ac{ps}-type of \acp{ef}, the propagation distances, \acp{aoa}, and \acp{aod} are given by $ d_{l,n} = \mathrm{d}_{\mathrm{ps}}(\V{p}_{n}, \V{p}_{l,\mathrm{mf}}, \V{p}_{\mathrm{pa}}) $, $ \varphi_{l,n} = \angle (\V{p}_{n}, \V{p}_{l,\mathrm{mf}}, \Delta\varphi_{n})$ and $ \phi_{l,n} = \angle (\V{p}_{l,\mathrm{mf}}, \V{p}_{\mathrm{pa}}, \Delta\phi)$.\footnote{$ \angle(\V{p}_{n}, \V{p}_{l,\mathrm{mf}}, \Delta\varphi_{n}) \triangleq \mathrm{atan2}\big(\frac{p_{l,\mathrm{mf},\mathrm{y}} - p_{n,\mathrm{y}}} {p_{l,\mathrm{mf},\mathrm{x}} - p_{n,\mathrm{x}}}\big) - \Delta\varphi_{n} $, $ \mathrm{d}_{\mathrm{va}} (\V{p}_{n}, \V{p}_{l,\mathrm{mf}}) \\ =  \|\V{p}_{n} - \V{p}_{l,\mathrm{mf}}\| $, and $ \mathrm{d}_{\mathrm{ps}}(\V{p}_{n}, \V{p}_{l,\mathrm{mf}}, \V{p}_{\mathrm{pa}}) = \|\V{p}_{n} - \V{p}_{l,\mathrm{mf}}\| + \|\V{p}_{l,\mathrm{mf}} - \V{p}_{\mathrm{pa}}\|  $.} To conveniently address the \ac{pa}-related variables and factors, we define $ \V{p}_{1,\mathrm{mf}} \triangleq \V{p}_{\mathrm{pa}} $.

\section{Radio Signal Model and Channel Estimation}
\label{sec:ProblemFromulation}

\subsection{Discrete-Frequency Signal Model}

The received signals are sampled with frequency spacing $ \Delta B $ over the bandwidth $ B $, yielding a length $ N_{\mathrm{f}} = B/\Delta B$ sample vector for each \ac{pa} and mobile agent antenna pair. By stacking the samples from all $ N_{\mathrm{rx}}N_{\mathrm{tx}} $ antenna pairs, we obtain the discrete-frequency signal vector $ \V{r}_{n} \rmv\in\rmv \mathbb{C}^{N_{\mathrm{rx}}N_{\mathrm{tx}}N_{\mathrm{f}}\rmv\times\rmv 1} $
\begin{align}
	\V{r}_{n} = \sum_{l=1}^{L_n} \V{B}(\V{\theta}_{l,n})\V{\alpha}_{l,n} + \V{n}_{n}
	\label{eq:SignalModel_freqDiscrete}\\[-7mm]\nn
\end{align}
where the first term comprises $ L_{n} $ \acp{mpc}, with each characterized by its state vector $ \V{\theta}_{l,n} \triangleq \big[ d_{l,n} \iist \phi_{l,n} \iist \varphi_{l,n} \big]^{\mathrm{T}}$ containing the delay $ \tau_{l,n}=d_{l,n}/c $, \ac{aod} $\phi_{l,n}$, \ac{aoa} $\varphi_{l,n}$, and complex amplitude $ \V{\alpha}_{l,n} \triangleq [\alpha_{\mathrm{hh},l,n} \iist \alpha_{\mathrm{hv},l,n} \iist \alpha_{\mathrm{vh},l,n} \iist \alpha_{\mathrm{vv},l,n}]^{\mathrm{T}}$.\footnote{The subscripts $ \{\mathrm{hh}, \mathrm{hv}, \mathrm{vh}, \mathrm{vv}\} $ denote four polarimetric transmission coefficients, e.g., $ \mathrm{hv} $ indexes the horizontal-to-vertical transmission coefficient.} We define the matrix $ \V{B}(\V{\theta}_{l,n}) \triangleq \big[\V{b}_{\mathrm{hh}}(\V{\theta}_{l,n}) \iist  \V{b}_{\mathrm{hv}}(\V{\theta}_{l,n}) \iist \V{b}_{\mathrm{vh}}(\V{\theta}_{l,n}) \iist  \V{b}_{\mathrm{vv}}(\V{\theta}_{l,n})\big] \rmv\in\rmv \mathbb{C}^{N_{\mathrm{rx}}N_{\mathrm{tx}}N_{\mathrm{f}}\rmv\times\rmv4} $ with columns given by $ \V{b}_{\mathrm{hv}}(\V{\theta}_{l,n}) \triangleq \mathrm{reshape} \big( \V{b}_{\mathrm{rx},\mathrm{h}}(\varphi_{l,n})\lozenge \V{b}_{\mathrm{tx},\mathrm{v}}(\phi_{l,n})\lozenge \V{b}_{\mathrm{f}}^{\mathrm{T}}(\tau_{l,n})\big) \rmv\in\rmv \mathbb{C}^{N_{\mathrm{rx}}N_{\mathrm{tx}}N_{\mathrm{f}}\rmv\times\rmv 1} $ and $ \lozenge $ denotes the Khatri–Rao product.\footnote{The operation $ \mathrm{reshape}(\cdot) $ reshapes a matrix into a column vector.} The vectors $ \V{b}_{\mathrm{rx},\mathrm{h}}(\varphi_{l,n}) \rmv\in\rmv \mathbb{C}^{N_{\mathrm{rx}}\rmv\times\rmv N_{\mathrm{f}}} $ and $ \V{b}_{\mathrm{tx},\mathrm{v}}(\phi_{l,n}) \rmv\in\rmv \mathbb{C}^{N_{\mathrm{tx}}\rmv\times\rmv N_{\mathrm{f}}} $ represent the far-field complex array responses by using the \ac{eadf} \cite{Xuesong_EADF2023}, and $ \V{b}_{\mathrm{f}}(\tau_{l,n}) \in \mathbb{C}^{N_{\mathrm{f}}\rmv\times\rmv 1} $ accounts for the system response, baseband signal spectrum and the phase shift due to delay $ \tau_{l,n} $ \cite{grebien2024SBL}. The measurement noise vector $ \V{n}_n \rmv\in\rmv \mathbb{C}^{N_{\mathrm{rx}}N_{\mathrm{tx}}N_{\mathrm{f}}\rmv\times\rmv 1} $ is a zero-mean, complex circular symmetric Gaussian random vector with covariance matrix $\V{C} = \sigma^2 \M{I}_{N_{\mathrm{rx}}N_{\mathrm{tx}}N_{\mathrm{f}}}$ where $\sigma^2$ is the noise variance. The \ac{mpc} \ac{snr} is given as the \ac{snr} calculated for $ \mathrm{hh} $ transmission, i.e., $ \mathrm{SNR}_{l,n} = \frac{|\alpha_{\mathrm{hh},l,n}|^2 \norm{\V{b}_{\mathrm{hh}}(\V{\theta}_{l,n})}^2} {\sigma^2}$ and the according normalized amplitude is $u_{l,n} = \sqrt{\mathrm{SNR}_{l,n}}$.

\subsection{Parametric Channel Estimation}

Based on the signal model in \eqref{eq:SignalModel_freqDiscrete}, a snapshot-based parametric channel estimation algorithm SAGE is applied in the pre-estimation stage\cite{SAGE_B_Fleury}, providing estimated dispersion parameters of $ M_{n} $ \acp{mpc} stacked into the vector $ \V{z}_{n} \rmv\triangleq\rmv [\V{z}_{1,n}^{\mathrm{T}} \ist \cdots \ist \V{z}_{M_n,n}^{\mathrm{T}}]^{\mathrm{T}} \rmv\in\rmv \mathbb{R}^{4M_{n}\rmv\times\rmv1}$. Each $ \V{z}_{m,n} \rmv\triangleq\rmv [{z_\mathrm{d}}_{m,n} \iist {z_\mathrm{\phi}}_{m,n} \iist {z_\mathrm{\varphi}}_{m,n} \iist {z_\mathrm{u}}_{m,n}]^{\mathrm{T}} $ comprises estimates ${z_\mathrm{d}}_{m,n}$ of the distance, the estimates ${z_\mathrm{\phi}}_{m,n}$ of the \ac{aod}, the estimates ${z_\mathrm{\varphi}}_{m,n}$ of the \ac{aoa}, and the estimates ${z_\mathrm{u}}_{m,n} \rmv\in\rmv [u_{\mathrm{de}}, \infty)$ of normalized amplitude, as well as of the noise variance. The estimates $ \V{z}_{n} $ above the detection threshold $ u_{\mathrm{de}} $ are used as noisy measurements by the proposed algorithm.

\section{System model}
\label{sec:SystemModel}

\subsection{Agent State and PMF States}
At each time $ n $, the state of mobile agent is given by $ \RV{x}_{n} \triangleq [\RV{p}_{n}^{\mathrm{T}} \iist \RV{v}_{n}^{\mathrm{T}}]^{\mathrm{T}}$ consisting of the position $ \RV{p}_{n} $ and the velocity $ \RV{v}_{n} =[\rv{v}_{\mathrm{x},n} \iist \rv{v}_{\mathrm{y},n}]^{\mathrm{T}} $. We assume that the array of the mobile agent is rigidly coupled with the movement direction, i.e., azimuth array orientation $ \rv{{\Delta\varphi}}_{n}$ is determined by the direction of its velocity vector $ \RV{v}_{n} $, i.e., $ {\Delta\varphi}_{n}(\V{v}_{n}) = \mathrm{atan2}(\frac{v_{\mathrm{y},n}}{v_{\mathrm{x},n}})$. All agent states up to time $ n $ are denoted as $ \RV{x}_{1:n} \triangleq [\RV{x}_{1}^{\mathrm{T}} \ist\cdots\ist \RV{x}_{n}^{\mathrm{T}} ]^{\mathrm{T}} $.

Following \cite{Erik_SLAM_TWC2019, Erik_DataFusion2023}, we account for the unknown and time-varying number of \acp{ef} by introducing \acp{pef} indexed by $k \in \{1,\dots, K_{n}\}$, where $ K_{n} $ represents the maximum possible number of \acp{ef} that produced a measurement so far and $ K_{n} $ increases with time. Augmented states of \acp{pef} are denoted as $ \RV{y}_{k,n} \triangleq [ \RV{{\mu}}_{k,n}^{\mathrm{T}} \iist \rv{r}_{k,n} ]^{\mathrm{T}} $ with $ \RV{{\mu}}_{k,n} = [\RV{p}^{\mathrm{T}}_{k,\mathrm{mf}} \iist \rv{u}_{k,n}]^{\mathrm{T}} $. The existence/non-existence of the $ k $th \ac{pef} is modeled by a binary random variable $ \rv{r}_{k,n} \in \{0,1\} $ in the sense that it exists if and only if $r_{k,n} = 1$. The type of the $ k $th \ac{pef} is modeled by a random variable $ \rv{q}_{k,n} \in \{1,2\} $ in the sense that the $ k $th \ac{pef} is a \ac{va}-type of \ac{ef} if $q_{k,n} = 1$ and it is a \ac{ps}-type of \ac{ef} if $q_{k,n} = 2$. Formally, \ac{pef} $ k $ is also considered even if it is non-existent, i.e., $r_{k,n} = 0$. The states $ \RV{{\mu}}_{k,n} $ of non-existent \acp{pef} are obviously irrelevant and have no influence on the \ac{pef} detection and state estimation. Therefore, all \acp{pdf} defined for \ac{pef} states $ f(\RV{y}_{k,n}) = f(\RV{{\mu}}_{k,n}, \rv{r}_{k,n}) $  are of the form $ f(\RV{{\mu}}_{k,n}, 0) = f_{k,n} f_{\mathrm{D}}(\RV{{\mu}}_{k,n}) $, where $ f_{\mathrm{D}}(\RV{{\mu}}_{k,n}) $ is an arbitrary ``dummy \ac{pdf}'' and $ f_{k,n} \in [0,1]$ is a constant representing the probability of nonexistence \cite{Florian_Proceeding2018, Florian_TSP2017, Erik_SLAM_TWC2019}. 

\subsection{Measurement Model}
\label{subsec:MeaModel}
Before the measurements are observed, they are considered as random and denoted as $ \RV{z}_{n} \rmv\triangleq\rmv [\RV{z}_{1,n}^{\mathrm{T}} \ist \cdots \ist \RV{z}_{M_n,n}^{\mathrm{T}}]^{\mathrm{T}} \rmv\in\rmv \mathbb{R}^{4\rv{M}_{n}\rmv\times\rmv1}$ and $ \RV{z}_{m,n} \rmv\triangleq\rmv [{\rv{z}_\mathrm{d}}_{m,n} \iist {\rv{z}_\mathrm{\phi}}_{m,n} \iist {\rv{z}_\mathrm{\varphi}}_{m,n} \iist {z_\mathrm{u}}_{m,n}]^{\mathrm{T}} $. An existing \ac{pef} generates a \ac{pef}-originated measurement $ \RV{z}_{m,n} $ with detection probability $ {p_{\mathrm{d}}}(\rv{u}_{k,n}) $ corresponding to the normalized amplitude $ \rv{u}_{k,n} $. The measurement \ac{lhf} is assumed to be conditionally independent across the individual measurements within the vector $ \RV{z}_{m,n} $. The individual \acp{lhf} of the distance, \ac{aoa} and \ac{aod} measurements are modeled by Gaussian \acp{pdf}. More specifically, the \acp{lhf} of the distance measurements for \ac{va}-originated and \ac{ps}-originated paths are given by 
\begin{align}
	& f_{q_{k,n} = 1}({z_\mathrm{d}}_{m,n} | \V{p}_{n}, \V{{\mu}}_{k,n}) \nn \\ 
	& \hspace{8mm} = f_{\mathrm{N}}({z_\mathrm{d}}_{m,n}; \mathrm{d}_{\mathrm{va}}(\V{p}_{n},\V{p}_{k,\mathrm{mf}}), \sigma_{\mathrm{d}}^2(u_{k,n})),  \label{eq:LHF_dist_VA} \\ 
	& f_{q_{k,n} = 2}({z_\mathrm{d}}_{m,n} | \V{p}_{n}, \V{{\mu}}_{k,n},\V{p}_{1,\mathrm{mf}}) \nn \\ 
	& \hspace{8mm} =  f_{\mathrm{N}}({z_\mathrm{d}}_{m,n}; \mathrm{d}_{\mathrm{ps}}(\V{p}_{n},\V{p}_{k,\mathrm{mf}},\V{{p}}_{1,\mathrm{mf}}), \sigma_{\mathrm{d}}^2(u_{k,n})),
	\label{eq:LHF_dist_PS} 
\end{align}
respectively. The \acp{lhf} of the \ac{aod} measurements for the \ac{los} path are given by
\begin{align}
	& f({z_\mathrm{\phi}}_{m,n} | \V{{p}}_{1,\mathrm{mf}}, \V{p}_{n}, u_{k,n}) \nn \\
	&\hspace{8mm} = f_{\mathrm{N}}({z_\mathrm{\phi}}_{m,n}; \angle(\V{{p}}_{1,\mathrm{mf}}, \V{p}_{n},\Delta\phi), \sigma_{\mathrm{\phi}}^2(u_{k,n}) ).
	\label{eq:LHF_aAoD_LOS}
\end{align}
The \acp{lhf} of the \ac{aod} measurements for \ac{ps}-originated path are given by
\begin{align}
	& f({z_\mathrm{\phi}}_{m,n} | \V{{p}}_{1,\mathrm{mf}}, \V{{\mu}}_{k,n}) \nn \\
	&\hspace{8mm} = f_{\mathrm{N}}({z_\mathrm{\phi}}_{m,n}; \angle(\V{{p}}_{1,\mathrm{mf}}, \V{p}_{k,\mathrm{mf}},\Delta\phi), \sigma_{\mathrm{\phi}}^2(u_{k,n}) ).
	\label{eq:LHF_aAoD}
\end{align}
The \acp{lhf} of the \ac{aoa} measurements are given by
\begin{align}
	& f({z_\mathrm{\varphi}}_{m,n} | \V{x}_{n}, \V{{\mu}}_{k,n}) \nn \\
	&\hspace{8mm} = f_{\mathrm{N}}({z_\mathrm{\varphi}}_{m,n}; \angle(\V{p}_{n},\V{p}_{k,\mathrm{mf}},\Delta\varphi(\V{v}_{n})), \sigma_{\mathrm{\varphi}}^2(u_{k,n}) ).
	\label{eq:LHF_aAoA}
\end{align}
The variances $ {\sigma_\mathrm{d}}^2(u_{k,n}) $, $ \sigma_{\mathrm{\phi}}^2(u_{k,n}) $ and $ \sigma_{\mathrm{\varphi}}^2(u_{k,n}) $ depend on the normalized amplitude $u_{k,n}$ and are determined based on the Fisher information given as $ {\sigma_\mathrm{d}}^2(u_{k,n}) = c^2/(8\pi^2 \beta_{\mathrm{bw}}^2 u_{k,n}^2) $, $ \sigma_{\phi_{k,n}}^2(u_{k,n}) = c^2/(8\pi^2f_{\mathrm{c}}^2 u_{k,n}^2 D^2(\phi_{k,n})) $ and $ \sigma_{\varphi_{k,n}}^2(u_{k,n}) = c^2/(8\pi^2f_{\mathrm{c}}^2 u_{k,n}^2 D^2(\varphi_{k,n})) $ with $ \beta_{\mathrm{bw}}^2$ denoting the mean square bandwidth of the transmit signal pulse and $ D^2(\cdot) $ is the squared array aperture \cite{Thomas_Asilomar2018, LeitingerICC2019}.\footnote{$ D^2(\phi_{k,n}) = \frac{1}{N_{\mathrm{tx}}}\sum_{i = 1}^{N_{\mathrm{tx}}} \frac{(d_{\mathrm{tx},i}\sin(\theta_{i})\sin(\phi_{i} -\phi_{k,n} ) )^2}{c^2} $, where $ d_{\mathrm{tx},i} $, $\phi_{i}$ and $ \theta_{i} $ denote the distance, azimuth and elevation angles from the transmit array center to the $ i $th antenna element. The squared array aperture for receive antenna array $ D^2(\varphi_{k,n}) $ is defined in the same way.}

The \acp{lhf} of the normalized amplitude measurements $ u_{k,n}\rmv>\rmv u_{\mathrm{de}} $ is modeled by a truncated Rician \ac{pdf} \cite[Ch.\,1.6.7]{BarShalom_AlgorithmHandbook}\cite{XuhongTWC2022}, i.e.,
\begin{align}
 \hspace*{-3mm} f({z_\mathrm{u}}_{m,n} | u_{k,n}) = f_{\mathrm{R}}({z_\mathrm{u}}_{m,n}; u_{k,n}, \sigma_{\mathrm{u}}(u_{k,n}), {p_{\mathrm{d}}}(u_{k,n}); u_{\mathrm{de}})
	\label{eq:LHF_normAmp}
\end{align}
where the scale parameter $ \sigma_{\mathrm{u}}(u_{k,n}) $ corresponding to $u_{k,n}$ is determined based on the Fisher information given as $ \sigma_{\mathrm{u}}^2(u_{k,n}) = \frac{1}{2} + \frac{1}{4N_{\mathrm{rx}}N_{\mathrm{tx}}N_{\mathrm{f}}} u_{k,n}^2 $. The detection probability $ {p_{\mathrm{d}}}(u_{k,n}) $ is modeled by the Marcum Q-function, i.e., $ {p_{\mathrm{d}}}(u_{k,n}) = Q_{1}(u_{k,n}/{\sigma_\mathrm{u}}_{k,n}, u_{\mathrm{de}}/{\sigma_\mathrm{u}}_{k,n}) $ \cite{BarShalom_AlgorithmHandbook,XuhongTWC2022}. 
 
Using \eqref{eq:LHF_dist_VA} to \eqref{eq:LHF_normAmp}, the \acp{lhf} $ f_{q_{k,n}}(\V{z}_{m,n}|\V{x}_{n}, \V{{\mu}}_{k,n}, \V{{\mu}}_{1,n}) $ for measurements originated from different type of \acp{pef} are given as follows.
\subsubsection{\ac{lhf} for \ac{va}-originated path} 
The \ac{lhf} for \ac{va}-originated paths is $ f_{q_{k,n} = 1}(\V{z}_{m,n}|\V{x}_{n}, \V{{\mu}}_{k,n}) $, which factorizes as
\begin{align}
	 f_{q_{k,n} = 1}(\V{z}_{m,n}|\V{x}_{n}, \V{{\mu}}_{k,n}) & = f_{q_{k,n} = 1}({z_\mathrm{d}}_{m,n} | \V{p}_{n}, \V{{\mu}}_{k,n}) \nn \\
	&\hspace{-15mm} \times  f({z_\mathrm{\varphi}}_{m,n} | \V{x}_{n}, \V{{\mu}}_{k,n})f({z_\mathrm{u}}_{m,n} | u_{k,n}).
	\label{eq:LHF_VA}
\end{align}
Note that the \ac{los} path is considered as a \ac{va}-originated path, but the corresponding \ac{lhf} also accounts for the \ac{aod} measurement, yields
\begin{align}
	&  f_{q_{k,n} = 1}(\V{z}_{m,n}|\V{x}_{n}, \V{{\mu}}_{1,n}) \nn \\
	& \hspace{2mm} = f_{q_{k,n} = 1}({z_\mathrm{d}}_{m,n} | \V{p}_{n}, \V{{\mu}}_{1,n}) f({z_\mathrm{\phi}}_{m,n} | \V{p}_{1,\mathrm{mf}}, \V{p}_{n}, u_{1,n}) \nn \\
	&\hspace{2mm}  \times  f({z_\mathrm{\varphi}}_{m,n} | \V{x}_{n}, \V{{\mu}}_{1,n})f({z_\mathrm{u}}_{m,n} | u_{1,n})
	\label{eq:LHF_LOS}
\end{align}
\subsubsection{\ac{lhf} for \ac{ps}-originated path}
\begin{align}
	& f_{q_{k,n} = 2}(\V{z}_{m,n}|\V{x}_{n}, \V{{\mu}}_{k,n}, \V{{\mu}}_{1,n}) \nn \\
	& = f_{q_{k,n} = 2}({z_\mathrm{d}}_{m,n} | \V{p}_{n}, \V{{\mu}}_{k,n}, \V{p}_{1,\mathrm{mf}})f({z_\mathrm{\phi}}_{m,n} | \V{p}_{n}, \V{{\mu}}_{k,n}) \nn \\
	&\hspace{2mm}  \times  f({z_\mathrm{\varphi}}_{m,n} | \V{x}_{n}, \V{{\mu}}_{k,n})f({z_\mathrm{u}}_{m,n} | u_{k,n})
	\label{eq:LHF_PS}
\end{align}

\subsubsection{\ac{lhf} for FAs}
We assume that the \ac{fa} measurements originating from the snapshot-based parametric channel estimator are statistically independent of \ac{pef} states. They are modeled by a Poisson point process with mean number $ \mu_{\mathrm{fa}} $ and \ac{pdf} $ f_{\mathrm{fa},q_{k,n}}(\V{z}_{m,n}) $. For \ac{va}-originated paths, the \ac{fa} \ac{pdf} is factorized as $ f_{\mathrm{fa},q_{k,n} = 1}(\V{z}_{m,n}) =  f_{\mathrm{fa}}({z_\mathrm{d}}_{m,n})f_{\mathrm{fa}}({z_\mathrm{\varphi}}_{m,n}) f_{\mathrm{fa}}({z_\mathrm{u}}_{m,n} )$. For \ac{los} path and \ac{ps}-originated paths, the \ac{pdf} is factorized as $ f_{\mathrm{fa},q_{k,n} = 2}(\V{z}_{m,n}) =  f_{\mathrm{fa}}({z_\mathrm{d}}_{m,n})f_{\mathrm{fa}}({z_\mathrm{\phi}}_{m,n})f_{\mathrm{fa}}({z_\mathrm{\varphi}}_{m,n}) f_{\mathrm{fa}}({z_\mathrm{u}}_{m,n} )$. The \acp{lhf} of \ac{fa} measurements corresponding to distance, azimuth angle and elevation angle are uniformly distributed on $ [0,d_{\mathrm{max}}] $, $ [-\pi,\pi) $ and $ [0,\pi] $, respectively. The false alarm \ac{lhf} $ f_{\mathrm{fa}}({z_\mathrm{u}}_{m,n} ) $ of the normalized amplitude is given by a truncated Rayleigh \ac{pdf} (see \cite{XuhongTWC2022, AlexTWC2024} for details).  
\subsection{State-Transition Model}
For each \ac{pef} with state $\RV{y}_{k,n-1} $ with $k \rmv\in\rmv \{1,\dots, K_{n-1}\}$ at time $n-1$, there is one ``legacy'' \ac{pef} with state $ \underline{\RV{y}}_{k,n} \rmv\rmv\triangleq\rmv\rmv [\underline{\RV{{\mu}}}_{k,n}^{\mathrm{T}} \iist \underline{\rv{r}}_{k,n}]^{\mathrm{T}} $ with $k \rmv\rmv\in\rmv\rmv \{1,\dots, K_{n-1}\}$ at time $n$. We define the stacked \ac{pef} state vector $ \underline{\RV{y}}_{n} \rmv\rmv\triangleq\rmv\rmv [\underline{\RV{y}}_{1,n}^{\mathrm{T}} \ist \cdots \ist  \underline{\RV{y}}_{K_{n-1},n}^{\mathrm{T}} ]^{\mathrm{T}} $ and stacked \ac{pef} type state vector $ \underline{\RV{q}}_{n} = [\underline{\rv{q}}_{1,n} \ist \cdots \ist \underline{\rv{q}}_{K_{n-1},n}]^{\mathrm{T}} $. Following the \ac{imm} approach, the temporal evolution of \ac{pef} type index $\rv{q}_{k,n}$ is modeled by a discrete Markov chain with constant transition matrix $\V{Q} \in [0,1]^{2\times2}$ over time, and the transition \ac{pmf} is given by $p(\underline{q}_{k,n}=1|q_{k,n}=2) = [\V{Q}]_{2,1} $ with $\sum_{i=1}^{2}[\V{Q}]_{i',i} = 1 \iist \forall \iist i' $ .\footnote{$[0,1]^{2\times2}$ denotes a $ 2\times2 $ matrix with entries between 0 and 1.} The \ac{pef} type index is assumed to evolve independently across $ k $ and $ n $, this the factorized prior \ac{pmf} of joint state is given as
\begin{align}
	p(\underline{\V{q}}_{n} | \V{q}_{n-1}) = \prod_{k=1}^{K_{n-1}} p(\underline{q}_{k,n}|q_{k,n-1})
	\label{eq:StateTransPDF_feaType} 
\end{align}
where $ p(\V{q}_{0}) $ is the initial \ac{pef} type \ac{pmf} at time $ n=0 $. The agent state and the legacy \acp{pef} states are assumed to evolve independently across $ k $ and $ n $ according to state-transition \acp{pdf} $ f(\V{x}_{n}|\V{x}_{n-1}) $ and $ f(\underline{\V{y}}_{k,n}|\V{y}_{k,n-1}) $, respectively, yields
\begin{align}
	f(\V{x}_{n}, \underline{\V{y}}_{n}|\V{x}_{n-1}, \V{y}_{n-1}) = f(\V{x}_{n}|\V{x}_{n-1}) \prod_{k=1}^{K_{n-1}} f(\underline{\V{y}}_{k,n}|\V{y}_{k,n-1})
	\label{eq:StateTransPDF_agentFea} \\[-8mm]\nn
\end{align}
where the formulation of the augmented \ac{pef} state-transition \ac{pdf} $ f(\underline{\V{y}}_{k,n}|\V{y}_{k,n-1}) = f(\underline{\V{\mu}}_{k,n}, \underline{r}_{k,n}|\V{\mu}_{k,n-1}, r_{k,n-1}) $ is inline with \cite{Florian_Proceeding2018,Erik_SLAM_TWC2019}.  

\subsection{New \acp{pef}}

Newly detected \acp{ef} at time $n$, i.e., \acp{pef} that generate measurements for the first time at time $n$, are modeled by a Poisson point process with mean $\mu_{\mathrm{n}}$ and \ac{pdf} $f_{\mathrm{n}}(\overline{\V{\mu}}_{m,n})$. Following \cite{Florian_Proceeding2018, Erik_SLAM_TWC2019}, newly detected \acp{pef} are represented by new \ac{pef} states $ \overline{\RV{y}}_{m,n} \triangleq [\overline{\RV{x}}_{m,n}^{\mathrm{T}}\iist \overline{\rv{r}}_{m,n}]^{\mathrm{T}} $, $ m \in \{1,\dots, \rv{M}_n\} $. Each new \ac{pef} $ \overline{\RV{y}}_{m,n} $ corresponds to a measurement $\RV{z}_{m,n}$, and $\overline{r}_{m,n} = 1$ means that the measurement $\RV{z}_{m,n}$ was generated by a newly detected \ac{pef}. The state vector of all new \acp{pef} at time $n$ is given by $ \overline{\RV{y}}_{n} \triangleq [\overline{\RV{y}}_{1,n}^{\mathrm{T}} \ist\cdots\ist \overline{\RV{y}}_{\rv{M}_{n},n}^{\mathrm{T}} ]^{\mathrm{T}} $ and the stacked \ac{pef} type state vector is given by $ \overline{\RV{q}}_{n} = [\overline{\rv{q}}_{1,n} \ist \cdots \ist \overline{\rv{q}}_{\rv{M}_{n},n}]^{\mathrm{T}} $. The new \acp{pef} $ \overline{\RV{y}}_{n} $ become legacy \acp{pef} at time $n+1$, accordingly the number of legacy \acp{pef} is updated as $K_{n} = K_{n-1} + M_{n}$. We also define $ \RV{y}_{n} \triangleq [\underline{\RV{y}}_{n}^{\mathrm{T}} \iist \overline{\RV{y}}_{n}^{\mathrm{T}} ]^{\mathrm{T}} $ with $\RV{y}_{k,n}$ and $ k \in \{1,\dots, K_{n}\} $, and $ \RV{q}_{n} \triangleq [\underline{\RV{q}}_{n}^{\mathrm{T}} \iist \overline{\RV{q}}_{n}^{\mathrm{T}} ]^{\mathrm{T}} $. The state and type vectors for all times up to $n$ are given by $ \RV{y}_{1:n} \triangleq [\RV{y}_{1}^{\mathrm{T}} \ist\cdots\ist \RV{y}_{n}^{\mathrm{T}} ]^{\mathrm{T}} $ and $ \RV{q}_{1:n} \triangleq [\RV{q}_{1}^{\mathrm{T}} \ist\cdots\ist \RV{q}_{n}^{\mathrm{T}} ]^{\mathrm{T}} $, respectively.

\subsection{Data Association}
Estimation of multiple \ac{pef} states is complicated by the \ac{da} uncertainty. Furthermore, it is not known if a measurement did not originate from a \ac{pef} (false alarm), or if a \ac{pef} did not generate any measurement (missed detection). The associations between measurements and legacy \acp{pef} are described by the \emph{\ac{pef}-oriented} association vector $ \RV{\underline{a}}_{n} \triangleq [\rv{\underline{a}}_{1,n} \ist \cdots \ist  \rv{\underline{a}}_{\rv{K}_{n-1},n}]^{\mathrm{T}} $ with entries $ \rv{\underline{a}}_{k,n} \triangleq m \rmv\in\rmv \{1,\dots, \rv{M}_n\}$, if legacy \ac{pef} $ k $ generates measurement $ m $, or $ \rv{\underline{a}}_{k,n} \triangleq 0 $, if legacy \ac{pef} $ k $ does not generate any measurement. In line with \cite{WilliamsLauTAE2014,Florian_Proceeding2018,Erik_SLAM_TWC2019}, the associations can be equivalently described by a \emph{measurement-oriented} association vector $ \RV{\overline{a}}_{n} \triangleq [\rv{\overline{a}}_{1,n} \ist \cdots \ist \rv{\overline{a}}_{\rv{M}_{n},n}]^{\mathrm{T}} $ with entries $ \rv{\overline{a}}_{m,n} \triangleq k \rmv\in\rmv \{1,\dots, K_{n-1}\} $, if measurement $ m $ is generated by legacy \ac{pef} $ k $, or $ \rv{\overline{a}}_{m,n} \triangleq 0 $, if measurement $ m $ is not generated by any legacy \ac{pef}. Furthermore, we assume that at any time $ n $, each \ac{pef} can generate at most one measurement, and each measurement can be generated by at most one \ac{pef} \cite{WilliamsLauTAE2014, Florian_Proceeding2018, Erik_SLAM_TWC2019}. The ``redundant formulation'' of using $ \RV{\underline{a}}_{n} $ together with $ \RV{\overline{a}}_{n} $ is the key to make the algorithm scalable for large numbers of \acp{pef} and measurements. The association vectors for all times up to $n$ are given by $\RV{\underline{a}}_{1:n} \triangleq [\RV{\underline{a}}_{1}^{\mathrm{T}} \ist\cdots\ist \RV{\underline{a}}_{n}^{\mathrm{T}} ]^{\mathrm{T}}$ and $ \RV{\overline{a}}_{1:n} \triangleq [\RV{\overline{a}}_{1}^{\mathrm{T}} \ist\cdots\ist \RV{\overline{a}}_{n}^{\mathrm{T}} ]^{\mathrm{T}} $.

\subsection{Joint Posterior PDF}
By using common assumptions \cite{BarShalom_AlgorithmHandbook, Florian_Proceeding2018, Erik_SLAM_TWC2019}, the joint posterior \ac{pdf} of $ \RV{x}_{0:n} $, $ \RV{y}_{0:n} $, $ \RV{q}_{1:n} $, $ \RV{\underline{a}}_{1:n} $ and $ \RV{\overline{a}}_{1:n} $ conditioned on observed (thus fixed) measurements $ \V{z}_{1:n} $ is given by
\begin{align}
	& \hspace*{0mm}f(\V{x}_{0:n}, \V{y}_{0:n}, \V{q}_{0:n}, \V{\underline{a}}_{1:n}, \V{\overline{a}}_{1:n} | \V{z}_{1:n}) \nonumber \\
	& \hspace*{-2mm}\propto \left(f(\V{x}_{0}) \prod_{i = 1}^{K_{0}} f(\V{y}_{i,0}) f(q_{i,0}) \right) \prod_{n' = 1}^{n} f(\V{x}_{n'}|\V{x}_{n'-1}) \nonumber \\
	& \hspace*{-2mm}\times \left( \prod_{k' = 1}^{K_{n'-1}} f(\underline{\V{y}}_{k',n'}|\V{y}_{k',n'-1}) p(\underline{q}_{k',n'}|q_{k',n'-1}) \right)  \nn \\
	& \hspace*{-2mm} \times \left(\prod_{k = 1}^{K_{n'-1}} g(\underline{\V{y}}_{k,n'}, \underline{q}_{k',n'}, \underline{a}_{k,n'}, \V{x}_{n'}; \V{z}_{n'}) \prod_{m = 1}^{M_{n'}} \psi(\underline{a}_{k,n'},\overline{a}_{m,n'})\right) \nn \\
	& \hspace*{-2mm} \times \left(\prod_{m' = 1}^{M_{n'}} h(\overline{\V{y}}_{m',n'}, \overline{q}_{m',n'}, \overline{a}_{m',n'}, \V{x}_{n'}; \V{z}_{m,n'})\right)
	\label{eq:jointPDF}
\end{align}
where the pseudo \acp{lhf} $ g(\underline{\V{y}}_{k,n}, \underline{q}_{k,n}, \underline{a}_{k,n}, \V{x}_{n}; \V{z}_{n}) = g(\underline{\V{\mu}}_{k,n}, \\ \underline{r}_{k,n}, \underline{q}_{k,n}, \underline{a}_{k,n}, \V{x}_{n}; \V{z}_{n}) $ related to legacy \acp{pef} are given by
\begin{align}
	& g(\underline{\V{\mu}}_{k,n}, \underline{r}_{k,n}=1, \underline{q}_{k,n}, \underline{a}_{k,n}, \V{x}_{n}; \V{z}_{n}) \nn \\
	& \hspace*{0mm} \triangleq
	\begin{cases}
		\dfrac{ f_{q_{k,n}}(\V{z}_{m,n}|\V{x}_{n}, \underline{\V{\mu}}_{k,n}, \underline{\V{\mu}}_{1,n})  p_{\mathrm{d}}(\underline{u}_{k,n}) } {\mu_{\mathrm{fa}} f_{\mathrm{fa},q_{k,n}}(\V{z}_{m,n})}, 										& \underline{a}_{k,n} = m \\
		1 - p_{\mathrm{d}}(\underline{u}_{k,n}),													 & \underline{a}_{k,n} = 0
	\end{cases}
	\label{eq:g} \\[-3mm]\nn
\end{align}
and $ g(\underline{\V{\mu}}_{k,n}, \underline{r}_{k,n}=0, \underline{q}_{k,n}, \underline{a}_{k,n}, \V{x}_{n}; \V{z}_{n}) \triangleq \delta(\underline{a}_{k,n})$. The pseudo \acp{lhf} $ h(\overline{\V{y}}_{m,n}, \overline{q}_{m,n}, \overline{a}_{m,n}, \V{x}_{n}; \V{z}_{m,n}) = h(\overline{\V{\mu}}_{m,n},\\ \overline{r}_{m,n}, \overline{q}_{m,n}, \overline{a}_{m,n}, \V{x}_{n}; \V{z}_{m,n}) $ related to new \acp{pef} are given by   
\begin{align}
	& h(\overline{\V{\mu}}_{m,n}, \overline{r}_{m,n}=1, \overline{q}_{m,n}, \overline{a}_{m,n}, \V{x}_{n}; \V{z}_{m,n}) \nn \\ 
	& \hspace*{-1mm} \triangleq
	\begin{cases} 
		0, 																				& \overline{a}_{m,n} = k \\
		\dfrac{ \mu_{\mathrm{n}} f_{\mathrm{n}}(\overline{\V{\mu}}_{m,n}) f_{q_{k,n}}(\V{z}_{m,n}|\V{x}_{n}, \overline{\V{\mu}}_{m,n}, \overline{\V{\mu}}_{1,n}) } { \mu_{\mathrm{fa}} f_{\mathrm{fa},q_{k,n}}(\V{z}_{m,n}) }, 								& \overline{a}_{m,n} = 0
	\end{cases}
	\label{eq:h} \\[-4mm]\nn
\end{align}
and $ h(\overline{\V{\mu}}_{m,n}, \overline{r}_{m,n}=0, \overline{q}_{m,n}, \overline{a}_{m,n}, \V{x}_{n}; \V{z}_{m,n}) \triangleq f_{\mathrm{D}}(\overline{\V{\mu}}_{k,n}) $. The detailed derivations of the joint posterior \ac{pdf}, the binary check function $ \psi(\underline{a}_{k,n}, \overline{a}_{m,n}) $ and the factor graph representation of \eqref{eq:jointPDF} are in parts inline with \cite{Florian_Proceeding2018, Erik_SLAM_TWC2019, XuhongTWC2022}.

\section{Problem Formulation and Proposed Method}
\label{sec:ProblemFormulation}

Using all the measurements up to time $ n $, we aim to sequentially detect \acp{pef} and estimate their positions and  and agent state. This relies on the marginal posterior existence probabilities $ p(r_{k,n} = 1 | \V{z}_{1:n}) $, the marginal posterior \acp{pdf} $ f(\V{p}_{k,\mathrm{mf}} | r_{k,n} = 1,\V{z}_{1:n}) $, $ f(q_{k,n} | r_{k,n} = 1,\V{z}_{1:n}) $ and $ f(\V{x}_{n}|\V{z}_{1:n}) $. More specifically, a \ac{pef} is detected if $ p(r_{k,n} = 1|\V{z}_{1:n}) \rmv>\rmv p_\mathrm{de} $, with $ p_\mathrm{de} $ denoting the detection threshold. The agent state $ \RV{x}_{n} $, and the states $\RV{p}_{k,\mathrm{mf}}$ and $\rv{q}_{k,n}$ of detected \acp{pef} are estimated by means of the \ac{mmse} estimator \cite{Kay_EstimationTheory}, i.e.,
\begin{align}	
	\hat{\V{x}}_{n} & \triangleq \int \V{x}_{n} f(\V{x}_{n}|\V{z}_{1:n})\mathrm{d} \V{x}_{n} \label{eq:MMSE_x} \vspace*{-2mm}\\
	\hat{\V{p}}_{k,\mathrm{mf}} & \triangleq \int \V{p}_{k,\mathrm{mf}} f(\V{p}_{k,\mathrm{mf}} | r_{k,n} = 1,\V{z}_{1:n}) \mathrm{d} \V{p}_{k,\mathrm{mf}}
	\label{eq:MMSE_pef} \vspace*{-2mm}\\
	\hat{q}_{k,n} & \triangleq \hspace*{-0mm} \sum_{i \in \V{Q} } \hspace*{-0mm} i f(q_{k,n} | \V{z}_{1:n})
	\label{eq:MMSE_featureType} 
	\vspace{-1mm}
\end{align}
where $ f(\V{p}_{k,\mathrm{mf}} | r_{k,n}=1,\V{z}_{1:n}) = \sum_{q_{k,\mathrm{mf}} \in \{1,2\}} f(\V{p}_{k,\mathrm{mf}}, q_{k,n} |\\  r_{k,n} = 1,  \V{z}_{1:n}) $
Since the marginal posterior \acp{pdf} $ f(\V{x}_{n}|\V{z}_{1:n}) $, $ p(r_{k,n} = 1 | \V{z}_{1:n}) $ and $ f(\V{p}_{k,\mathrm{mf}} | r_{k,n} = 1,\V{z}_{1:n}) $ cannot be obtained analytically, we use a computationally efficient sequential particle-based message-passing implementation by means of sum-product algorithm rules to obtain approximations of these marginal posterior \acp{pdf}. As the number of \acp{pef} grows with time $n$ (at each time by $ K_{n} = K_{n-1} + M_{n}$), \acp{pef} with $ p(r_{k,n} = 1 | \V{z}_{1:n})$ below a threshold $p_\mathrm{pr}$ are removed from the state space (``pruned'').

\section{Experimental Results}
\label{sec:ExperimentalResults}

The performance of the proposed algorithm is validated using both synthetic and real radio measurements, for which the following setup and parameters are commonly used. 

\subsection{Analysis Setup}
The state-transition \ac{pdf} $f(\V{x}_{n}|\V{x}_{n-1}) $ of the agent is defined by a linear near constant-velocity motion model\cite{BarShalom_AlgorithmHandbook}, given as $ \V{x}_{n} = \V{F}\V{x}_{n-1} + \V{\Gamma}\V{\nu}_{n} $, where the matrix $\V{F} \in \mathbb{R}^{4\times4} $ and $ \V{\Gamma} \in \mathbb{R}^{4\times2} $ are chosen as in \cite{BarShalom_AlgorithmHandbook} with the sampling period $ \Delta T $. The driving process $ \V{\nu}_{n} \in \mathbb{R}^{2\times1} $ is iid across time $ n $, zero-mean and Gaussian with covariance matrix $ \sigma_{\mathrm{\nu}}^2\mathbf{I}_{2} $, $ \mathbf{I}_{2} $ denotes a $ 2\rmv\times\rmv2 $ diagonal matrix and $ \sigma_{\mathrm{\nu}}^2 = 0.0025\,$$ \text{m}/\text{s}^2 $ represents the average speed increment along $ x $ or $ y $ axis during $ \Delta T $. The state-transition \ac{pdf} of legacy \acp{pef} $\underline{\V{p}}_{k,\mathrm{mf}}$ is chosen to be $ \underline{\V{p}}_{k,\mathrm{mf}} = \V{p}_{k,\mathrm{mf}} + {\V{\epsilon}}_{k,n} $, where the noise ${\V{\epsilon}}_{k,n}$ is iid across $ k $ and $ n $, zero-mean, and Gaussian with variance ${\underline{\sigma}_{\mathrm{p}}}_{k}^2\mathbf{I}_{2}$ and $ {\underline{\sigma}_{\mathrm{p}}}_{k} = 10^{-5}\,$m. The state-transition \ac{pdf} of the normalized amplitude $\underline{\rv{u}}_{k,n}$ is chosen to be $ \underline{u}_{k,n} = u_{k,n-1} + {\epsilon_{\mathrm{u}}}_{k,n} $, where the noise ${\epsilon_{\mathrm{u}}}_{k,n}$ is iid across $ k $ and $ n $, zero-mean, and Gaussian with variance $\sigma_{\mathrm{u}}^2(\underline{u}_{k,n})$. The \ac{pef} mode transition probabilities are chosen as $ [\V{Q}]_{1,1} = [\V{Q}]_{2,2}  = 0.96 $ and $ [\V{Q}]_{1,2} = [\V{Q}]_{2,1}  = 0.04 $.

We assume that the geometric environment information is not available as a prior. The samples for the initial agent state are drawn from a $ 4 $-D uniform distribution centered at $ [\V{p}_{0}^{\mathrm{T}}\iist 0 \iist 0]^{\mathrm{T}} $ where $ \V{p}_{0} $ is the true agent start position, and the support of position and velocity components are $ [-0.2,0.2]\,$m and $ [0.02,0.02]\,$m/s, respectively. The samples for the initial states $ \V{{\mu}}_{k,n} $ of a new \ac{pef} are drawn from a $ 4 $-D Gaussia distribution with means $ [{z_\mathrm{d}}_{m,n} \iist {z_\mathrm{\phi}}_{m,n} \iist {z_\mathrm{\varphi}}_{m,n} \iist {z_\mathrm{u}}_{m,n}]^{\mathrm{T}} $ and variances calculated using the amplitude measurements $ {z_\mathrm{u}}_{m,n} $ (see Section~\ref{subsec:MeaModel}). The mean number of new \acp{pef} is $\mu_{\mathrm{n}} = 0.1$, the probability of survival is $p_{\mathrm{s}} = 0.999$, the detection and pruning threshold are $ p_\mathrm{de} =0.5 $ and $p_\mathrm{pr} = 10^{-3}$, and the particle number is $ 200000 $. 

\subsection{Synthetic Measurement Evaluation}
\label{subsec:synMeaIntro}
\begin{figure}[t!]
	\centering
	\hspace*{6mm}\subfloat[\label{subfig:modeBelief_MF1}]
	{\hspace*{-9mm}\includegraphics[]{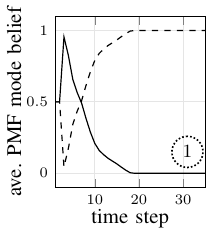}}
	\hspace*{2.5mm}\subfloat[\label{subfig:modeBelief_MF2}]
	{\hspace*{-2.5mm}\includegraphics[]{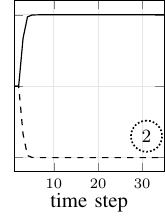}}
	\hspace*{1mm}\subfloat[\label{subfig:modeBelief_MF3}]
	{\hspace*{-1mm}\includegraphics[]{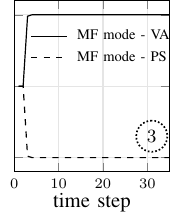}}
	\caption{Results for synthetic measurements. Averaged \ac{pef} mode beliefs associated to the three true \acp{ef} highlighted in Fig.~\ref{subfig:mappingResult}.}	 
	\label{fig:modeBelief}
	\vspace*{0mm}
\end{figure}

First, we present the simulation results using fully synthetic measurements without involving the snapshot-based channel estimator. We synthesized $ 3 $ \acp{mpc} with time-varying distances, angles and amplitudes according to the true agent positions over $ 100 $ time steps obtained in the real measurement and three true \acp{ef} (one \ac{ps} and two \acp{va}) highlighted with dotted circle markers in Fig.~\ref{subfig:mappingResult}. Fig.~\ref{fig:modeBelief} shows the two \ac{pef} mode beliefs of the three detected \acp{ef}, averaged over $ 20 $ simulation runs. It can be seen that the first detected \ac{ef} (associated with the highlighted \ac{ps} in Fig.~\ref{subfig:mappingResult}) rapidly converged to  the \ac{ps} mode, and the other two detected \acp{ef} (associated with the highlighted \acp{va} in Fig.~\ref{subfig:mappingResult}) rapidly converged to the \ac{va} mode. 

\subsection{Real Measurement Evaluation}

\begin{figure}[t!]
	\centering
	\includegraphics[width=8.4cm,height=4.6cm]{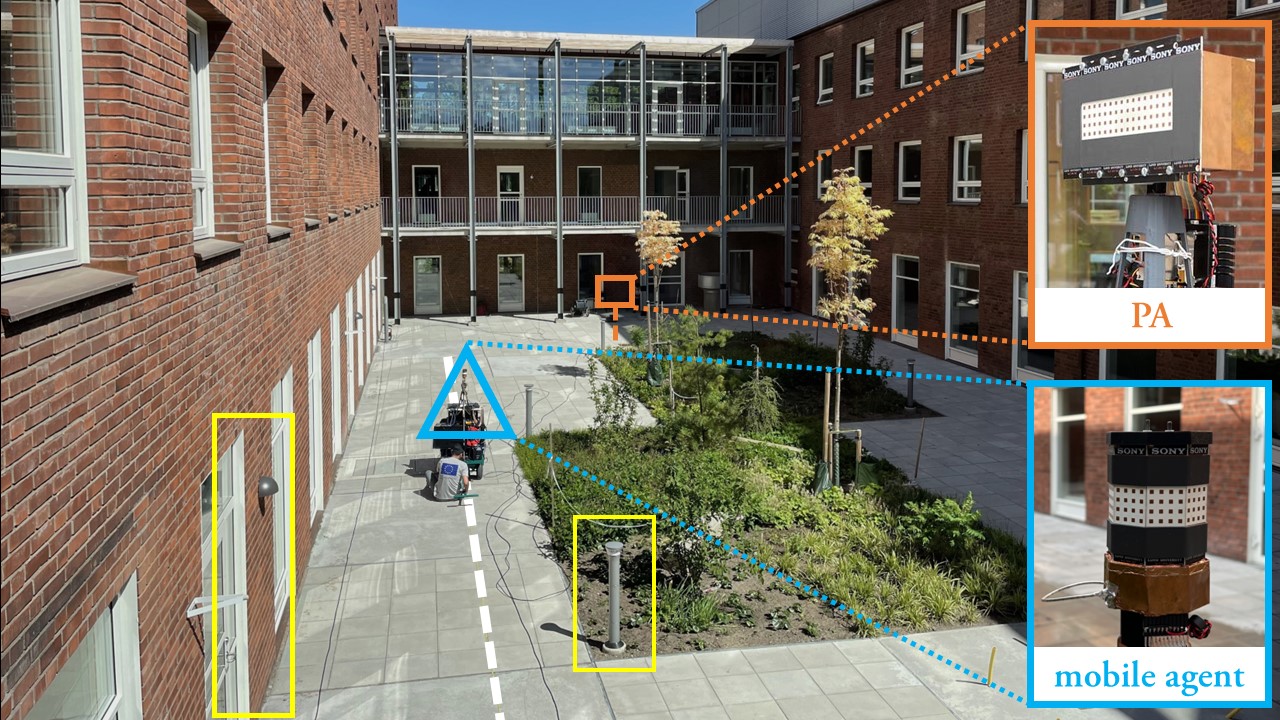}
	\caption{Overview of the measurement environment, Lund Sweden. The groud truth trajectory of the mobile agent is denoted with white dashed line. The static \ac{pa} position and the current mobile agent position are indicated by a orange square and a cyan triangle, the same markers are also applied in Fig.~\ref{fig:GraphicalOverview} and Fig.~\ref{subfig:mappingResult}. A metallic pillar and a window corner are highlighted by yellow squares as examples of distinct point scatterers. }
	\label{fig:MeaEnv}
	\vspace{0mm}
\end{figure}

\begin{figure*}[t]
	\hspace*{0mm}
	\begin{minipage}[t]{.32\textwidth}
		\vspace*{-75mm}
		\hspace*{6.5mm}\subfloat[\label{subfig:agentPosError}]
		{\hspace*{-7.5mm}\includegraphics[]{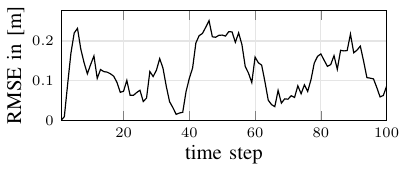}}\vspace*{-4mm} \\
		\hspace*{6mm}\subfloat[\label{subfig:agentOrientationError}]
		{\hspace*{-6.5mm}\includegraphics[]{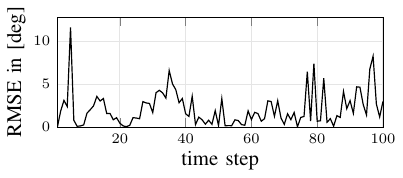}}
	\end{minipage}
	\hspace*{8mm}
	\begin{minipage}[t]{.65\textwidth}
		\hspace*{2.5mm}\subfloat[\label{subfig:mappingResult}]
		{\hspace*{-5mm}\includegraphics[]{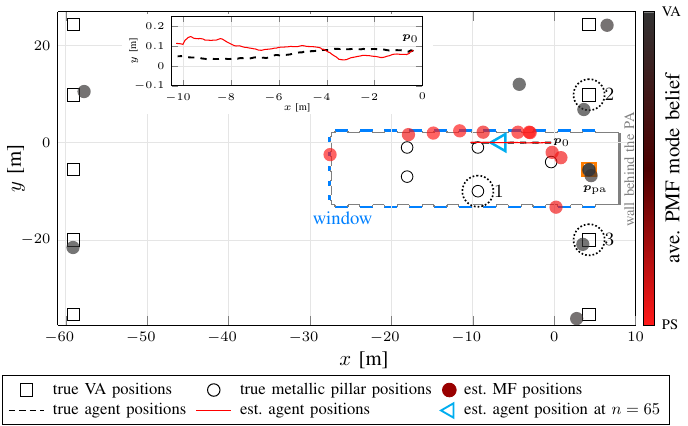}}
	\end{minipage}
	\caption{Performance results using real \ac{mmwave} massive \ac{mimo} measurements. (a) \acp{rmse} versus time of the agent position and (b) \acp{rmse} versus time of the agent orientation. (c) shows the floorplan of the measurement environment (Fig.~\ref{fig:MeaEnv}) including the windows, walls and true \ac{ef} positions. The zoomed-in plot on the top shows the true and estimated agent positions. The \ac{mmse} estimates of detected \ac{ef} positions at time $ n=65 $ are denoted as circle markers with the respective marker color representing the average of the two \ac{ef} mode probability estimates. The three true \acp{ef} (one \ac{ps} and two \acp{va}) highlighted by dotted circles were used for generating the synthetic measurements in Section~\ref{subsec:synMeaIntro}.}	 
	\label{fig:ResultsRealMea}
	\vspace{-0mm}
\end{figure*}

\subsubsection{Measurement Setup}

The performance is further validated using real \ac{mmwave} massive \ac{mimo} measurements collected in a countyard at Lund University, Sweden, as shown in Fig.~\ref{fig:MeaEnv}. The courtyard has approximate dimension of $ 35 $\,m$ \times $$ 15 $\,m$ \times $$ 13 $\,m, which presents a rich-scattering environment featuring vegetation and surrounded by brick walls with multiple reflective windows. The measurement used a switched array channel sounder supporting an effective bandwidth of $ 768 $\,MHz centered at $ 28 $\,GHz. On the \ac{pa} side, a uniform planar array with $ 64 $ dual-polarized patch antennas ($ 128 $ ports in total) was placed at a fixed known position with the main radiation direction facing the yard. At the mobile agent side, a cylindrical array with $ 128 $ dual-polarized patch antennas ($ 256 $ ports in total) was used and manually moved along a $ 10\,$m straight line trajectory. The channel impulse response was recorded every $ 10\,$cm, generating a total of $ 100 $ measurement snapshots. For some snapshots, \ac{los} propagation path are obstructed by the vegetation. The complex gain over both polarizations of antenna arrays were characterized in an anechoic chamber. More details on the \ac{mmwave} channel sounder can be found in \cite{Xuesong_mmWave2023}.  The ground truth of agent positions was obtained with a \ac{slam} system fusing measurements from a LiDAR sensor and an IMU sensor mounted on the cart holding the mobile agent array \cite{HediehEuCAP2024}. Considering the 2D formulation of the agent and \ac{ef} states, we only used the measurements from SAGE with elevation \acp{aoa} that are within $ 8 \,$degrees of the horizontal.


\subsubsection{Performance}
Fig.~\ref{subfig:agentPosError} shows the \acfp{rmse} of the agent positions versus time $ n $ which are mostly below $0.2\,$m, and the mean \ac{rmse} over the whole track is $0.12\,$m.   Fig.~\ref{subfig:agentOrientationError} shows the \acp{rmse} of the agent orientation versus time $ n $, which rapidly converge below $ 5\,$ degrees after $ 10 $ steps, and the mean orientation \ac{rmse} over the whole track is $ 2\,$ degrees. For an exemplary simulation run, Fig.~\ref{subfig:mappingResult} shows the estimated agent track and the estimated \acp{pef} at time $ n=65 $ with the marker color indicating the average belief of the two \ac{ef} modes. Given that \ac{pa} planar array was orientated towards the yard, no distinct \acp{mpc} (i.e., \acp{ef}) associated with the wall behind the \ac{pa} are detected. It is shown that the window corners along side the agent track, the window corners close to the $ 3 $rd highlighted \ac{ef}, and the metallic pillar close to the \ac{pa} are clearly detected with \ac{ps} as the dominant \ac{ef} mode. Furthermore, several \acp{va} up to the $ 2 $nd order are also detected and match the geometrical predicted \acp{va}. Note that in this environment, the signals scattered from the window corners act like a radar returns and are much stronger than the signals scattered from ``classical'' \acp{ps} such as metal pillars.



%

\vspace*{0mm}
\section{Conclusions}
\label{sec:Conclusions}

We presented a multipath-based \ac{slam} algorithm that continuously adapts interacting \ac{ef} models describing \acp{mpc} originating from specular reflection and point scattering. The interacting \ac{ef} model evolves over time according to a discrete Markov chain, which is incorporated into the factor graph representing the \ac{slam} problem. The results using real measurements demonstrate the great potential of \ac{mmwave} massive \ac{mimo} systems for accurate and robust localization in real and challenging outdoor scenarios, and the exceptional environment sensing capability of the proposed algorithm, compared to \ac{va}-only based methods. Possible directions for future research include extending the proposed algorithm to three-dimensional scenarios with horizontal and vertical
propagation, and introducing \acp{aod} to \ac{va}-related \acp{lhf}.



\section{Acknowledgment}
This work was supported in part by the Vinnova/FFI project Beyond 5G positioning under Grant 2022-01640, in part by the Strategic Research Area Excellence Center at Link\"oping--Lund in Information Technology (ELLIIT), in part by the Horizon Europe Framework Programme under the Marie Sk{\l}odowska-Curie grant agreement No. 101059091, in part by the Swedish Research Council (Grant No. 2022-04691), in part by the Royal Physiographic Society of Lund, in part by the Christian Doppler Research Association, and in part by the TU Graz. The authors thank Juan Sanchez, Hedieh Khosravi, and Christian Nelson for helping with the measurements.

\vspace*{10mm}
\renewcommand{\baselinestretch}{1}  
\bibliographystyle{IEEEtran}
\bibliography{IEEEabrv,./references}

\end{document}